\documentclass[a4paper]{jpconf}
\usepackage{graphicx}
\begin{document}
\title{Unveiling stellar magnetic activity using CoRoT seismic observations}

\author{Savita Mathur$^{1}$, Rafael A. Garc\'ia$^{2}$, David Salabert$^{3,4}$, J\'er\^ome Ballot$^{5}$, Clara R\'egulo$^{3,4}$, Travis S. Metcalfe$^{1}$, and Annie Baglin$^{6}$}

\address{$^1$ High Altitude Observatory, 3080 Center Green Drive, Boulder, CO, 80301, USA}
\address{$^2$ Laboratoire AIM, CEA/DSM-CNRS-Universit\'e Paris Diderot; CEA, IRFU, SAp, F-91191, Gif-sur-Yvette, France}
\address{$^3$ Instituto de Astrof\'isica de Canarias (IAC), E-38200 La Laguna, Tenerife, Spain}
\address{$^4$ Dept. de Astrof\'isica, Universidad de La Laguna (ULL), E-38206 La Laguna, Tenerife, Spain}
\address{$^5$ Laboratoire dÕAstrophysique de Toulouse-Tarbes, Universit\'e de Toulouse, CNRS, F-31400, Toulouse, France}
\address{$^6$ LESIA, UniversitŽ Pierre et Marie Curie, Universit\'e Denis Diderot, Observatoire de Paris, 92195 Meudon Cedex, France}

\ead{savita@ucar.edu}

\begin{abstract}
It is well known that in the Sun, the frequencies and amplitudes of acoustic modes vary throughout the solar cycle. Indeed, while the magnetic activity goes towards its maximum, the frequencies of the modes increase and their amplitudes decrease.
We have analyzed data from the CoRoT mission on a few stars that exhibit solar-like oscillations. The study of HD49933 (observed during 60 days and 137 days spanning a total of 400 days) showed a modulation of the maximum amplitude per radial mode and the frequency shifts of the modes, showing magnetic activity in this rapidly rotating star. Moreover, both properties vary in an anticorrelated way and the data allowed us to establish a lower limit for the activity-cycle period of ~120 days. Measurements in Ca H and K lines confirmed that this star is in the ``active stars'' category.
We will also discuss the results obtained for other targets such as HD181420 and HD49835 for which we have investigated a similar behavior. 
\end{abstract}

\section{Introduction}

Many improvements have been made on the mechanisms behind the solar activity cycle, which can be seen in the simulations of the solar dynamo [e.~g. 1]. However, the last solar minimum that was unexpectedly long emphasized the fact that our knowledge is not completely accurate and that classical proxies do not say the same thing as seismic parameters [2, 3, 4, 5, 6]. The recent discovery of a 2-year modulation in the seismic observables [7] raises more questions to the understanding of such cycles in the Sun.

One way to better understand the solar cycle is to study magnetic activity cycles in other stars. Baliunas et al. [8] showed that stellar dynamos depend on the evolutionary time scale of the star. There is already an empirical law that has been drawn out thanks to the observations of a few tens of stars [9, 10]. For cool stars like the Sun and with an $\alpha \Omega$ dynamo, a longer rotation period implies a longer cycle period. But this law depends utterly on the characteristics of the convective zone, which can be well determined thanks to asteroseismology. We know that the activity of the Sun interacts with the acoustic modes and it might also have an impact on the high-frequency p modes [11, 12].

The last decade,  asteroseismology made a big jump. Thanks to the CoRoT mission, several solar-like stars could be observed continuously during a few months with a good signal-to-noise ratio (SNR) [e.g.  13, 14, 15] or with a much lower one [16] allowing us to measure individual low-degree p modes in some of the former cases. 
Here we show how asteroseismology can allow us to detect a magnetic activity cycle in the Sun and other stars.

\section{Methodology}

For this analysis, we studied the temporal variations of the power spectrum, starspots number, frequency shifts of the p modes, and maximum amplitude per radial mode.

\subsection{Wavelet analysis}

We obtain an estimate of the surface rotation of the star with the wavelets techniques [17,18]. It was shown in [12, 19] that this tool is very powerful for helio- and asteroseismology. In particular, it distinguishes between the fundamental period and the first harmonic of the rotation period as in the classical power spectrum, we have this ambiguity sometimes. The wavelet power spectrum also shows the temporal evolution of the magnetic activity of the star.

\subsection{Proxy of the ``starspots number"}

We can also notice that the time series have the fingerprints of the starspots and the bright plages appearing and moving across the stellar disk. By calculating the standard deviation of small subseries of the data, we are able to follow the evolution of the global coverage of the starspots on the stellar surface [20]. This is a rough ``proxy'' for the magnetic activity cycle of a star as we cannot resolve the stellar disk. 

\subsection{Calculating the frequency shifts of the p modes}

We also studied the variation of the frequency shift of the global modes or the position of the p-mode hump. We have used two different techniques to obtain these frequency shifts.

The first one consists of studying the global envelope of the p-mode hump. We compute the power spectrum density of subseries  and subtract the background, modeled with three components (Harvey-law model, a power law, and white noise) and six free parameters (ignoring the p-mode region). The cross-correlation function [21]  is computed in the region of the p modes. These cross-correlation functions look like Gaussians. To determine a possible shift in this cross-correlation function, we use a frequency range of $\pm$~7~$\mu$Hz to estimate the third order moment of this function, which measures the asymmetry. Then, starting with the lag given by the asymmetry, we fit a Gaussian function using a window of $\pm$~ 7~$\mu$Hz. The maximum of this fitted Gaussian is used as the position of the cross-correlation peak.

We can also measure the shift of individual modes. Here, we fit the modes of each subseries with a standard likelihood maximization function [22]. We fit the modes with a Lorentzian profile over frequency windows containing the l=0, 1, and 2 modes. Then we compute the difference of frequency compared to a reference. Finally, the frequency shift is the average of the differences over the frequency range studied.

\subsection{Estimation of the maximum amplitude per radial mode}

We use the A2Z pipeline [19] to calculate the rms maximum amplitude per radial mode, A$_{max}$ and its variation with time. We subtract the background fit and smooth the power spectrum density over 2$\times \Delta\nu$, where $\Delta\nu$ is the mean large separation of the modes of the star. We fit the envelope of the modes with a Gaussian and convert the maximum power to bolometric amplitude per radial mode [23].

We have successfully applied our methodology to the Sun [20] using 10 years of data from the VIRGO (Variability of solar IRradiance and Gravity Oscillations [24]) instrument aboard the SoHO spacecraft,  taking the average of the three independent SPM (Sun PhotoMeters) channels (red, blue, and green). 

\section{Analysis of the star HD49933}

We analysed the data obtained by the CoRoT (Convection Rotation and planetary Transits) satellite for the star HD49933, which is an F5V star of 1.2~M$_\odot$ and 1.3~R$_\odot$. Around 50 acoustic modes have been identified thanks to the 60+137 days of CoRoT observations [25,26].
\begin{figure}[h]
\center
\includegraphics[width=10cm]{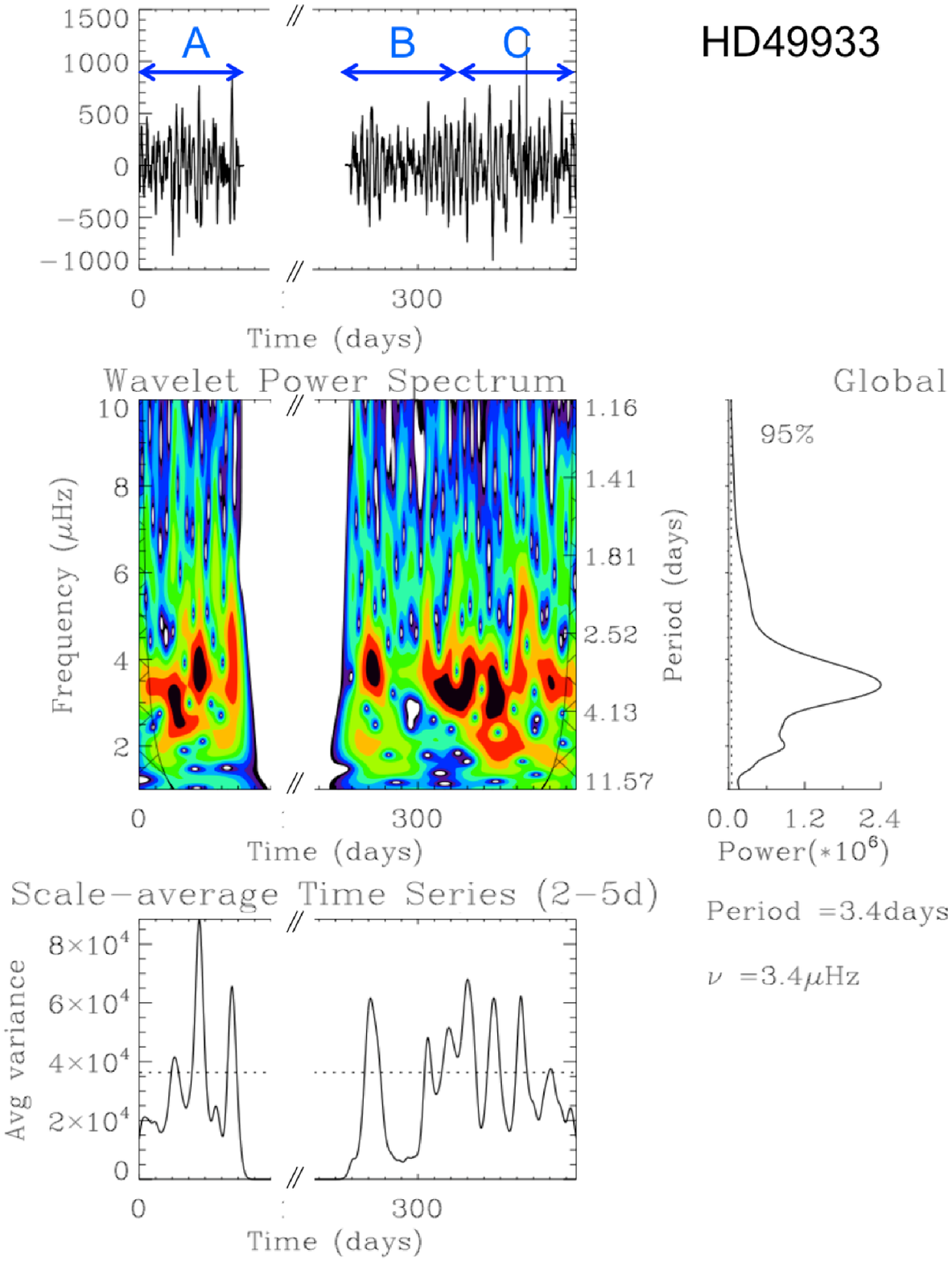}\hspace{2pc}%
\begin{minipage}[b]{10pc}
\caption{\label{fig2}Wavelet analysis of HD49933. Top panel: time series. Middle panel: wavelet power spectrum (left) and global wavelet power spectrum (right). Bottom panel: scale average variance between a period of 2 and 5 days.}
\end{minipage}
\end{figure}


Fig.~\ref{fig2} (top) shows the light curve of HD49933 observed during 60 days of the CoRoT initial run in 2007 and again during 137 days in 2008. It presents clear signatures of active regions. 
A surface rotation is confirmed at ~3.5~d with some differential rotation between 2 and 5 days (middle panel of Fig. 2).
The collapsogram between 2 and 5 days (bottom panel of Fig. 2) shows an increase in the average signal in the first run, A, while the first part of the long run, B, is a more quiet period. Finally, C is a more active period.

\begin{figure}[h]
\center
\includegraphics[width=8cm]{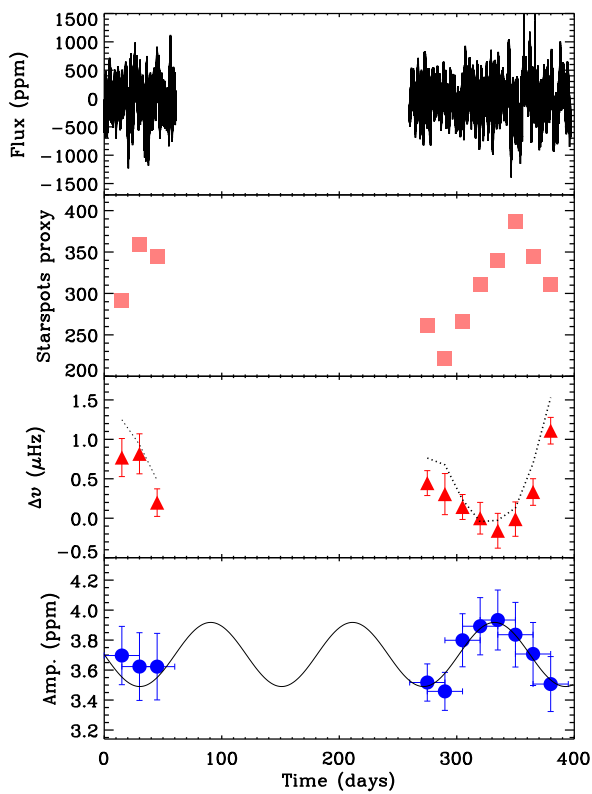}\hspace{2pc}%
\begin{minipage}[b]{14pc}
\caption{\label{fig3}Top: lightcurve of HD49933 obtained with CoRoT. Second panel: temporal evolution of the starspots proxy obtained as explained in Sect. 2.1. Third panel: temporal evolution of the frequency shifts ($\Delta\nu$) using the cross-correlation (red triangles) and the individual mode fitting (dotted line). Bottom: maximum amplitude per radial mode versus time. A tentative fit with a sinus wave is overplotted giving a lower limit for the cycle period of 120 days (see text for details).}
\end{minipage}
\end{figure}

We computed the starspots proxy as described in Section 2.3 leading to Fig.~\ref{fig3} (2nd panel from the top). The minimum of this proxy occurs at around 300~d.

We analyzed the parameters of the acoustic modes. We took subseries of 30 days shifted every 15 days. We calculated the cross-correlation of the p-mode hump (between 1460 and 2100~$\mu$Hz) of each subseries with a reference one. The latter is taken during the minimum of activity. It gives the red triangles of Fig.~\ref{fig3} (3rd panel). 
The frequency shifts of the individual modes were calculated in the range 1460-2070~$\mu$Hz. We removed the outliers above 5~$\sigma$. The frequency shifts obtained are the black curve in Fig.~\ref{fig3} (3rd panel).
After subtracting the background, we fitted the envelope of the modes with a Gaussian function to obtain the rms maximum amplitude per radial mode (Fig.~\ref{fig3}, bottom panel).

We clearly see the anticorrelation in the temporal evolution between the amplitude variation and the frequency shifts variation, confirming the existence of a magnetic activity cycle, which also agrees with the wavelet analysis. We fitted the whole set of data with a sinus wave. The best fit is obtained for a period of 120 days, which is a lower limit of the cycle period (the sinusoide is overplotted to guide the eye). Indeed, we do not have a full coverage of the cycle and we know from the Sun that the periods of maximum activity can have a ``plateau'' region that sometimes has some structure such as a double maximum. Because our fit is obtained around the minimum activity we should be cautious about the real length of the cycle and even about the fact that is a regular cycle.

Besides, we see a small time lag between the minimum of activity in the starspots proxy while in the asteroseismic data, the minimum comes 30 days later. This could be explained with the inclination angle from which we observe the star as it influences other seismic parameters as the rotational splitting [27, 28].  Some simulations show that depending on the angle of inclination, the minimum (or maximum) of activity is shifted in time and with more or less strength when we look for the signature of the starspots [29], This is a consequence of the interaction between the active longitudes, the migration of the spots towards the equator when the cycle evolves and the inclination of the star.

Some results were obtained thanks to observations of the Calcium H and K on April 13, 2010, showing that this is an active star with a Mount Wilson S-index of 0.3 [20] confirming previous conclusions [30]. Additional Ca H and K observations are scheduled for this fall. 

\section{Other CoRoT targets}
\begin{figure}[h]
\begin{minipage}{12pc}
\includegraphics[width=10pc, angle=90]{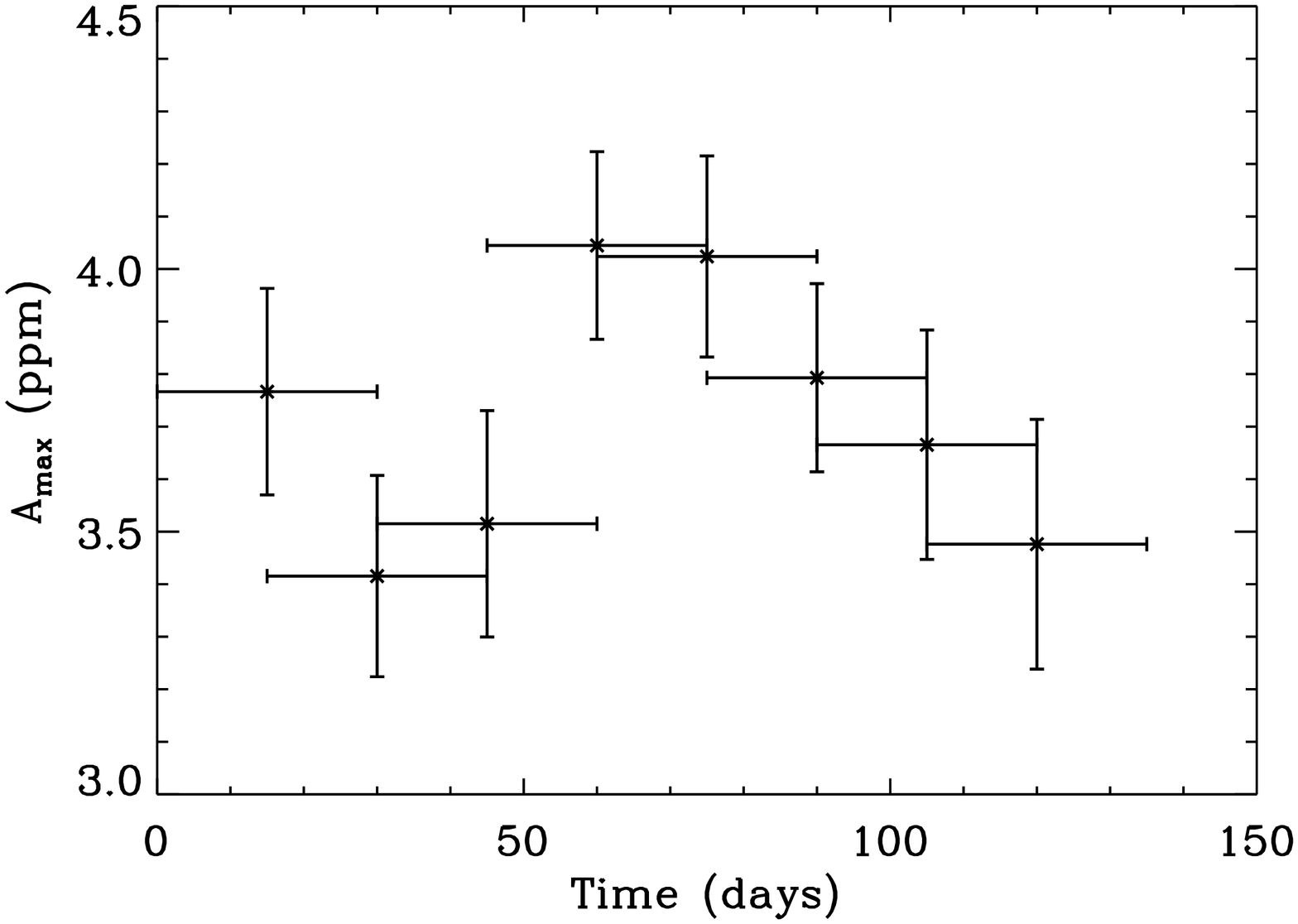}
\includegraphics[width=10pc, angle=90]{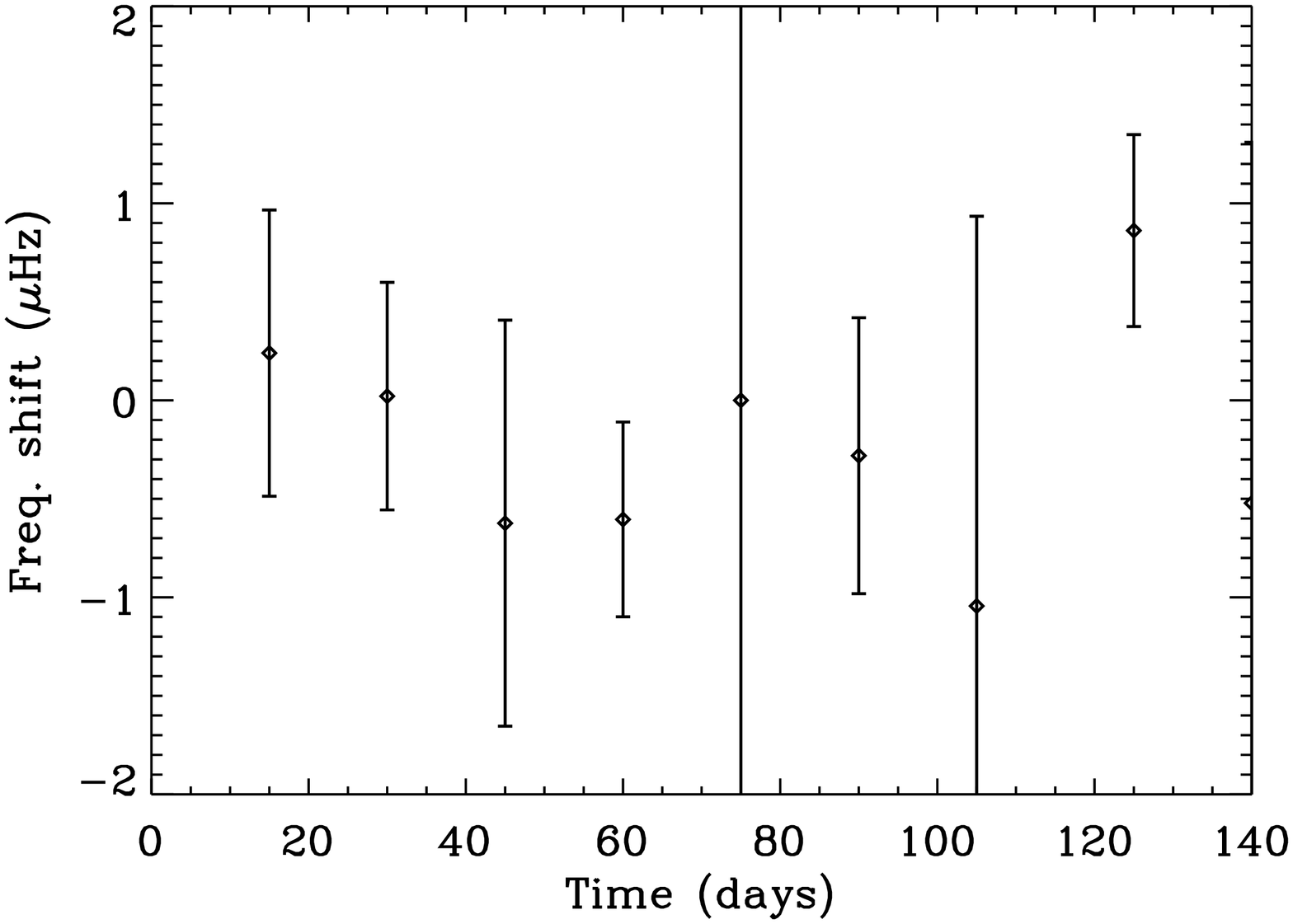}
\caption{\label{fig4}Amplitude and frequency shift variation for HD181420.}
\end{minipage}\hspace{2pc}%
\begin{minipage}{12pc}
\includegraphics[width=10pc, angle=90]{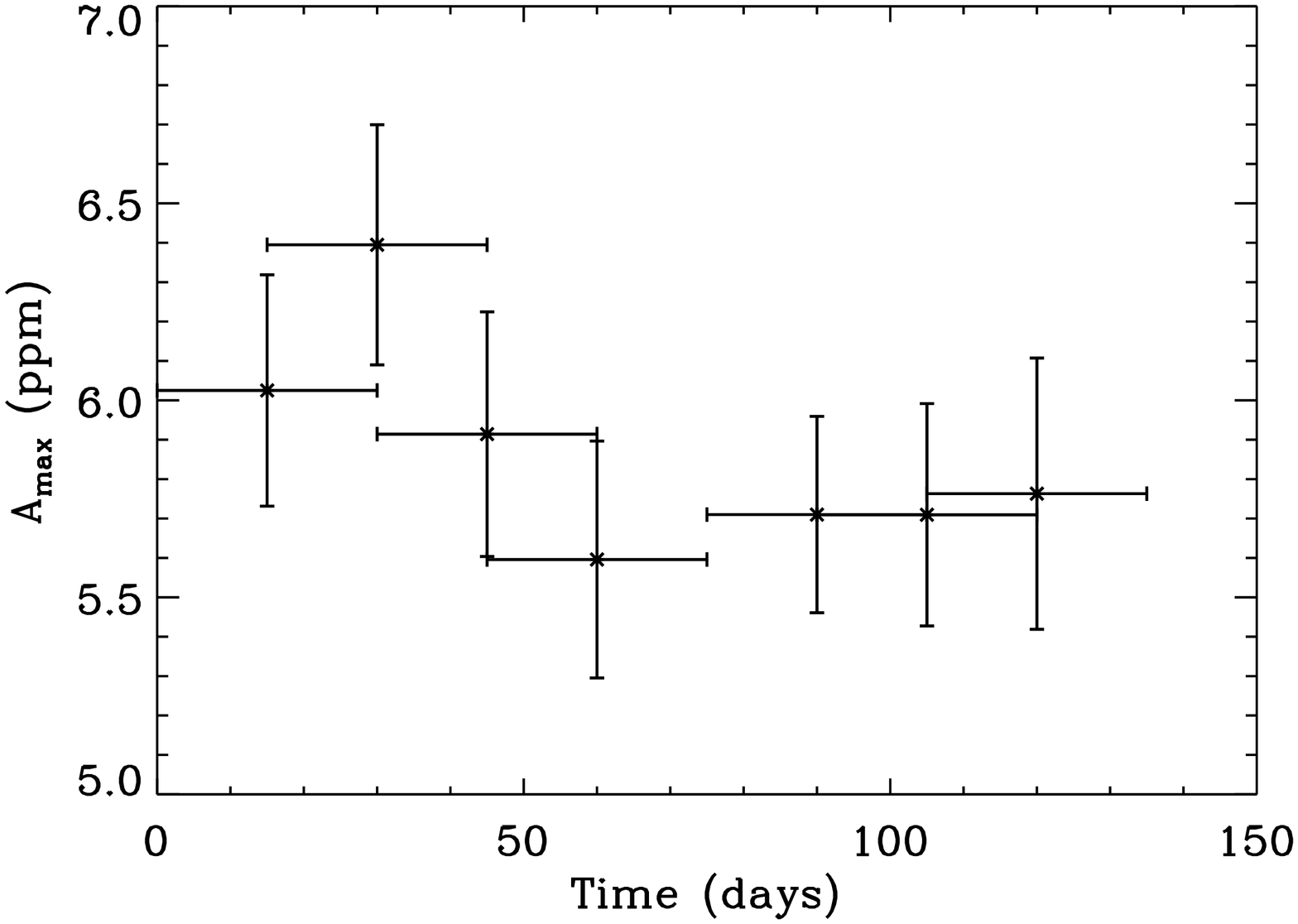}
\includegraphics[width=10pc, angle=90]{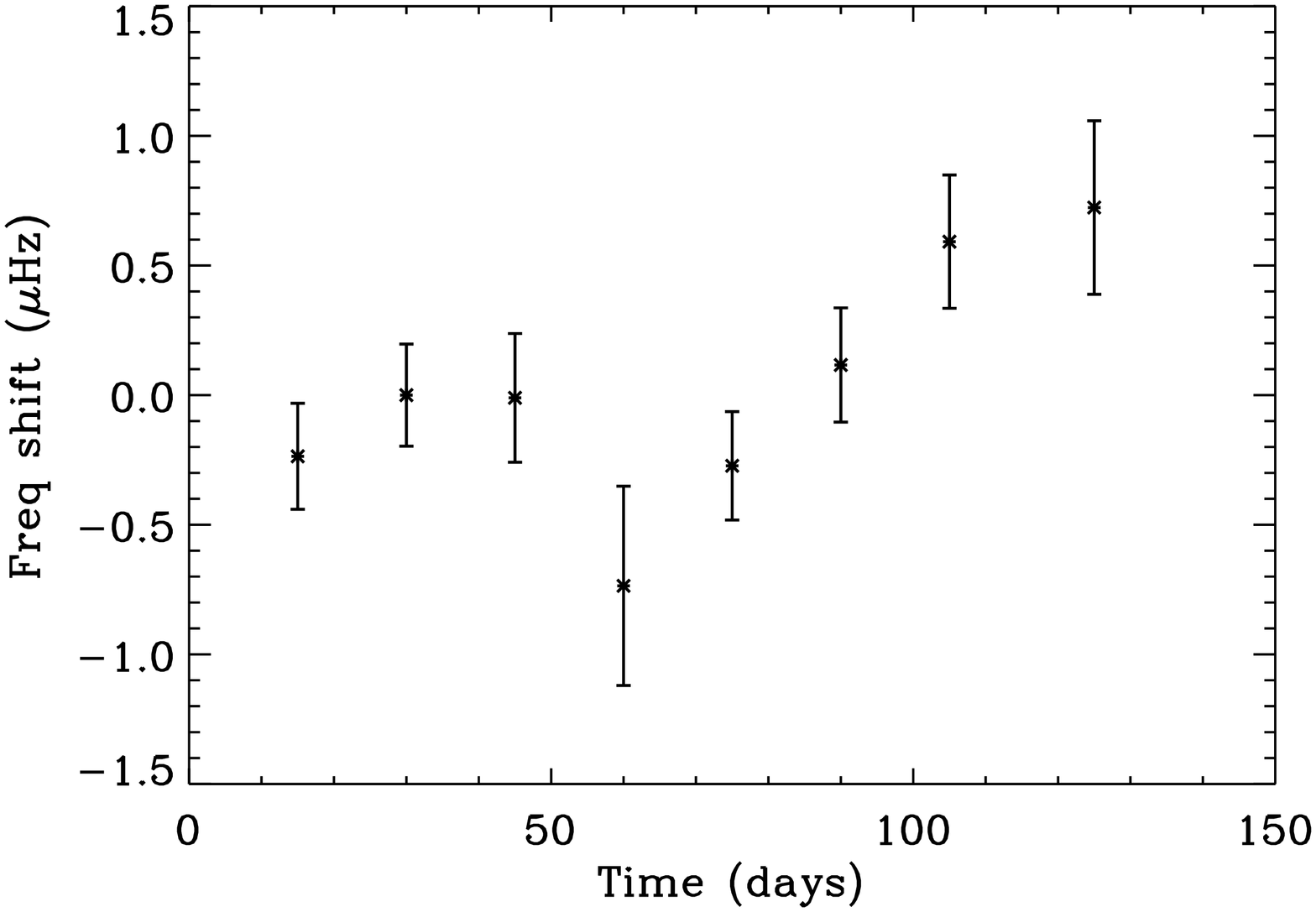}
\caption{\label{fig5}Amplitude and frequency shift variation for HD49385.}
\end{minipage} 
\end{figure}

We also applied the same methodology to other stars observed by CoRoT. Figs.~\ref{fig4} and \ref{fig5} show the temporal variations of the amplitude and the frequency shifts for HD181420 [13] and HD49385 [14]. We know that the first star has a rotation period of 2.6~d, while the rotation period of HD49385 is still uncertain, maybe around 10 days. For HD181420, there is a small hint of anticorrelation between the two parameters with a maximum for the amplitude around 60~d. But we cannot confirm it because of the large error bars, specially for the frequency shifts. For HD49385,  we do not see any anticorrelation, which could be due to the fact that for such a rotation period, the cycle might be much longer than the observation length.

\section{Conclusions}

These analyses show that HD49933 has a magnetic activity and we obtain a lower limit on the cycle period of ~120 days. Actually, such a short cycle might not be so uncommon as  observations in Ca HK of HD17051 show a cycle period of 1.6 yr  [31]. 
The long-term observations of Kepler will allow us to apply this methodology on more stars and give new constraints for stellar dynamo models. Thanks to the accuracy we can have on the convective zone parameters and on the modes parameters, it would enable us to better constrain the simulations. The fact that such short cycles can exist tells us that we should be able to observe these magnetic activity cycles during the 3.5 years of observations by the Kepler mission that is observing solar-like oscillating stars [32, 33, 34, 35, 36, 37].

\ack
NCAR is supported by the National Science Fundation. This work has been partially supported by the Spanish National Research Plan (grant PNAyA2007-62650) and by the ÒProgramme National de Physique StellaireÓ at CEA Saclay. 
The CoRoT space mission, launched on December 27th 2006, has been developed and is operated by CNES, with the contribution of Austria, Belgium, Brazil, ESA (RSSD and Science Program), Germany and  Spain.


\end{document}